\documentclass[pra,twocolumn,aps,superscriptaddress,nofootinbib]{revtex4-1}
\usepackage{amsfonts}
\usepackage{mathtools}
\usepackage[svgnames,table]{xcolor}
\usepackage[T1]{fontenc}
\usepackage{txfonts}
\usepackage{graphicx}
\usepackage{lineno}

\usepackage{hyperref}
\hypersetup{
        colorlinks=true,
        citecolor=SteelBlue,
        filecolor=LimeGreen,
        linkcolor=SlateBlue,
        urlcolor=MediumPurple
}

\usepackage{dcolumn}
\usepackage{bm}
\usepackage{amsmath}
\usepackage{breqn}
\usepackage[bbgreekl]{mathbbol}
\usepackage{bbm}
\usepackage{mathrsfs}
\usepackage[permil]{overpic}
\usepackage[labelfont=bf]{subcaption}
\usepackage{textgreek}

\newcommand{\comment}[2]{\textcolor{orange}{}}

\makeatletter
\let\cat@comma@active\@empty
\makeatother

\begin{document}
\preprint{APS/123-QED}
\title{Compact Michelson interferometers with subpicometer sensitivity}
\author{Jiri Smetana}\email{gsmetana@star.sr.bham.ac.uk}
\affiliation{Institute for Gravitational Wave Astronomy, School of Physics and Astronomy, University of Birmingham, Birmingham B15 2TT, United Kingdom}
\author{Rebecca Walters}
\affiliation{Institute for Gravitational Wave Astronomy, School of Physics and Astronomy, University of Birmingham, Birmingham B15 2TT, United Kingdom}
\author{Sophie Bauchinger}
\affiliation{Institute for Gravitational Wave Astronomy, School of Physics and Astronomy, University of Birmingham, Birmingham B15 2TT, United Kingdom}
\author{Amit Singh Ubhi}
\affiliation{Institute for Gravitational Wave Astronomy, School of Physics and Astronomy, University of Birmingham, Birmingham B15 2TT, United Kingdom}
\author{Sam Cooper}
\affiliation{Institute for Gravitational Wave Astronomy, School of Physics and Astronomy, University of Birmingham, Birmingham B15 2TT, United Kingdom}
\author{David Hoyland}
\affiliation{Institute for Gravitational Wave Astronomy, School of Physics and Astronomy, University of Birmingham, Birmingham B15 2TT, United Kingdom}
\author{Richard Abbott}
\affiliation{LIGO Laboratory, California Institute of Technology, Pasadena, CA 91125, USA}
\author{Christoph Baune}
\affiliation{SmarAct Metrology GmbH \& Co. KG, Rohdenweg 4, 26135 Oldenburg, Germany}
\author{Peter Fritchel}
\affiliation{LIGO Laboratory, Massachusetts Institute of Technology, Cambridge, Massachusetts 02139, USA}
\author{Oliver Gerberding}
\affiliation{Institute of Experimental Physics, University of Hamburg, Luruper Chaussee 149,
D-22761 Hamburg, Germany}
\author{Semjon K\"ohnke}
\affiliation{SmarAct Metrology GmbH \& Co. KG, Rohdenweg 4, 26135 Oldenburg, Germany}
\author{Haixing Miao}
\affiliation{Department of Physics, Tsinghua University, Beijing, China 100084}
\affiliation{Institute for Gravitational Wave Astronomy, School of Physics and Astronomy, University of Birmingham, Birmingham B15 2TT, United Kingdom}
\author{Sebastian Rode}
\affiliation{SmarAct Metrology GmbH \& Co. KG, Rohdenweg 4, 26135 Oldenburg, Germany}
\author{Denis Martynov}
\affiliation{Institute for Gravitational Wave Astronomy, School of Physics and Astronomy, University of Birmingham, Birmingham B15 2TT, United Kingdom}
\date{\today}

\begin{abstract}
The network of interferometric gravitational-wave observatories has successfully detected tens of astrophysical signals since 2015. In this paper, we experimentally investigate compact sensors that have the potential to improve the sensitivity of gravitational-wave detectors to intermediate-mass black holes. We use only commercial components, such as sensing heads and lasers, to assemble the setup and demonstrate its subpicometer precision. The setup consists of a pair of Michelson interferometers that use deep frequency modulation techniques to obtain a linear, relative displacement readout over multiple interference fringes. We implement a laser-frequency stabilisation scheme to achieve a sensitivity of 0.3\,$\text{pm} / \sqrt{\text{Hz}}$ above 0.1\,Hz. The device has also the potential to improve other experiments, such as torsion balances and commercial seismometers.
\end{abstract}

\maketitle

\section{\label{sec:intro}Introduction}

Interferometric displacement readout has played a pivotal role in expanding the boundaries of precision measurement. In the field of gravitational wave (GW) astronomy, interferometric readout had shown its utility in detecting GW signals~\cite{BBHDetection,Catalogue1,Catalogue2}. The cornerstone of its success, in part, lies in two areas: (a) the exceptional sensitivity achievable over distances significantly smaller than the wavelength of light used and (b) versatility in design across vastly different length scales. To the first point, contemporary GW detectors such as \textit{Advanced LIGO}~\cite{AdvLIGO} and \textit{Advanced Virgo}~\cite{AdvVirgo} can measure some of the weakest signals in the universe in the form of GW strain. In particular, \textit{Advanced LIGO} is capable of measuring displacements as fine as $2\times10^{-20} \text{m} / \sqrt{\text{Hz}}$ in the most sensitive frequency region around 100\,Hz~\cite{Buikema2020}. To the latter point, interferometric measurement sees widespread use in investigations of quantum phenomena but also spans km-scales in terrestrial GW detection and million km scales in \textit{LISA}---the future space-based GW observatory~\cite{LISA}.

We present the investigation of a compact, interferometeric sensor for the precise measurement of picometer-scale displacements of macroscopic objects. Displacement sensing of gram-scale optics through to kilogram-scale test masses is widespread in the GW community due to the strict requirements on suppression of terrestrial displacement noise (such as seismic noise) in ground-based GW detectors. Displacement sensing and control has been achieved through numerous means, such as capacitive position sensors~\cite{AdvLIGO}, coil-readout geophones such as the Sercel L-4C and optical shadow sensors such as the BOSEM~\cite{Bosems}. The displacement sensitivity of these devices is limited across a broad band by the readout noise of these sensors. When used in feedback control, the sensitivity of these instruments imposes a fundamental floor on the level of suppression that is achievable. For example, \textit{Advanced LIGO} is limited in its sensitivity below 30\,Hz by controls noise~\cite{Buikema2020}, which can be improved with more sensitive devices. The astrophysical case for expanding sensitivity towards lower frequency in contemporary GW detectors is broad, ranging from improved early-warning systems~\cite{EarlyWarning} for multi-messenger detections~\cite{MultiMessenger} to the routine observation of heavier astronomical objects such as intermediate-mass black holes~\cite{IMBHDetection}. 

The future role of interferometric displacement sensors in achieving the sensitivity improvements necessary for observing these astrophysical phenomena has already been explored in detail in~\cite{5Hz}. The 5--30\,Hz frequency band in \textit{Advanced LIGO} is dominated by a range of technical noises, such as angular controls noise caused by low frequency ($<1$\,Hz) ground vibrations (see~\cite{Buikema2020, 5Hz} and references therein).
Inertial sensors with an interferometeric readout, such as 6D seismometers~\cite{Compact6D} and next generation beam rotation sensors~\cite{BRS_2014, TorsionMichelson}, have the potential to significantly reduce the test mass motion, simplify the lock acquisition process~\cite{Staley_2014}, and reduce the required bandwidth of angular control loops. Furthermore, interferometeric sensors can efficiently damp the suspensions without limiting the design sensitivity of GW detectors. The sensors achieve the required improvement of sensing noise by a factor of 100 compared to the currently utilised shadow sensors~\cite{Bosems, 5Hz}.

Outside the field of GW detection, compact interferometry has seen use across a wide range of applications. It has been applied in the readout of torsion balance experiments~\cite{TorsionMichelson}, which can be used in investigations into areas of fundamental physics, such as short-range force measurements or bosonic dark matter searches~\cite{TorsionDarkMatter}. Compact interferometers have also been used as reference sensors in the development of other displacement measuring devices~\cite{CompactReferenceSensor}. The potential for compact interferometers to serve as absolute distance measurement devices~\cite{AbsoluteDistanceSensor} has led to their use in alignment sensing at the \textit{Large Hadron Collider}~\cite{LHCAlignment}.

The history of compact interferometer development reaches back several decades and includes a broad range of designs, topologies and varying levels of sophistication based on the technology available at the time. A comprehensive review of compact interferometry can be found in \cite{CompactReview} and expounds all of the different strategies undertaken in this sphere. Compact interferometry is not a recent concept as exemplified by the device by de la Rue et al. \cite{Compact1972} dating from 1972. However, the design of a high performance compact interferometer is non-trivial and often falls into a particular performance niche. The design can be tuned for high linearity, in one case 5\,pm, which was achieved by Weichert et al. \cite{CompactLinear}. Other designs are tuned for high sensitivity within the desired frequency range, such as the \textit{LISA Pathfinder} interferometer, achieving a sensitivity of $1 \text{pm} / \sqrt{\text{Hz}}$ at 10\,mHz \cite{CompactLISAPF}.

\begin{figure*}[t]
\centering
\includegraphics[width=0.98\textwidth]{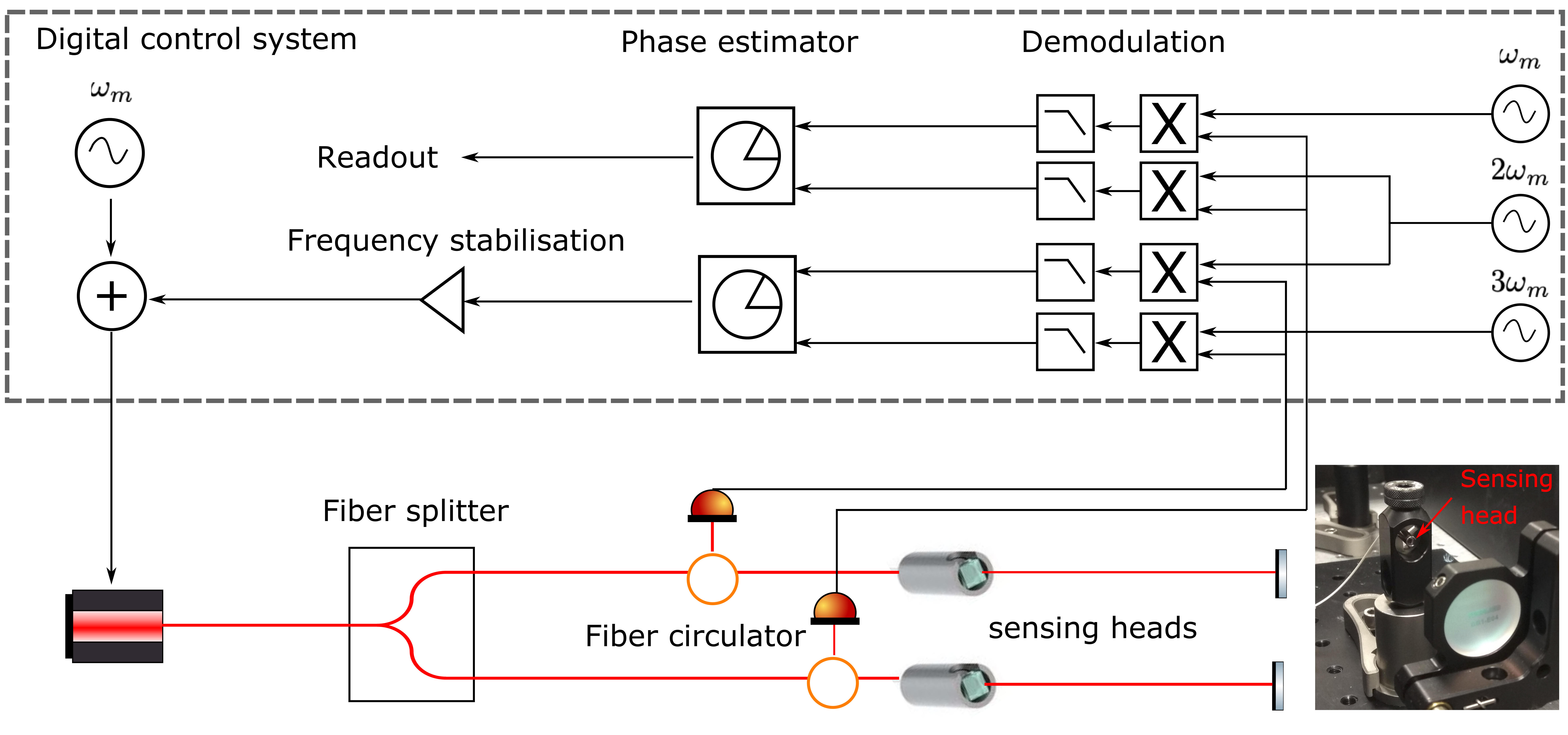}
\caption{Layout of the experiment and of the digital signal processing. A digitally sourced modulation is applied to the laser, which is then connected to both sensing heads. The raw DC diode signal is digitised and demodulated in the digital domain at multiple harmonics. An algorithm is applied to extract the displacement from the resulting signals.}
\label{fig:layout}
\end{figure*}

The term `compact' in reference to these devices can, however, be rather misleading depending on the application. Many of the existing compact interferometers still occupy an area of hundreds of square-centimeters, which is unacceptably large for applications where space is at a premium, such as the sensing of the \textit{Advanced LIGO} quadruple suspension~\cite{LIGOQUAD}. Some devices conform to the label of `compact' better than others, for example the design by Royer et al.~\cite{CompactSmall} is confined to an area of only $8 \times 5 \, \text{cm}$. Further development has led to interferometers designed with the direct application to seismic displacement sensing in mind. An example of this, the HoQI sensor \cite{HoQITest}, has shown substantial improvements in the sensitivity of commercial geophones by replacing the electronic coil readout and thus eliminating the dominant source of noise~\cite{HoQIGeophone}.

The high sensitivity is in part due to the very sharp response of the interferometer, being able to cover the full signal range in a mere quarter-wavelength of displacement. A typical value of 266\,nm for a simple Michelson interferomter, is significantly smaller than that achievable by alternate technology, such as the 0.7\,mm linear range of the BOSEM~\cite{Bosems}. The cost of this sharp response is a significant reduction in range, limiting the use of many interferometric devices to specialised uses over sub-wavelength range. However, we can implement multi-quadrature readout techniques~\cite{Isleif_2019}, coupled with a fringe-counting algorithm, to expand the range over multiple interference fringes (read \cite{CompactReview} for an overview of the many multi-fringe readout techniques available). These techniques maintain the precision of the instrument, whilst overcoming its most significant limitation.

In this paper, we build a compact displacement sensor out of commercially available products, such SmarAct C01 sensing heads, NKT Adjustik laser, and Thorlabs fiber beam splitters. We implement a multi-quadrature readout scheme with fringe-counting in the digital domain, which can be readily compared to the analogue-domain equivalent demonstrated by numerous other multi-fringe interferometers. The scheme makes use of a purely digital implementation of the laser frequency modulation, subsequent signal demodulation, filtering and arm phase extraction. The key strengths of this design are that the physical system is highly compact, mechanically simple, robust and easy to deploy where the necessary computational infrastructure is already present. We use only commercial components, with the primary component, the sensing head, occupying an area of merely $1.3 \times 0.4 \, \text{cm}$ (excluding any necessary mounting brackets). We achieve a high level of sensitivity due to an appropriate choice of laser and through additional frequency stabilisation. In Section~\ref{sec:theory} we briefly outline the principles of deep frequency modulation and present the experimental layout. Section~\ref{sec:sens} is devoted to the experimental results of our fixed-mirror measurement with the accompanying noise budget and the subsequent improvements we achieved using a frequency-stabilising feedback system.

\section{\label{sec:theory}Optical layout}

\begin{figure*}[t]
\begin{subfigure}{0.48\linewidth}
\begin{flushleft}\textbf{(a)}\end{flushleft}
\vspace{-3mm}
\includegraphics[width=\linewidth]{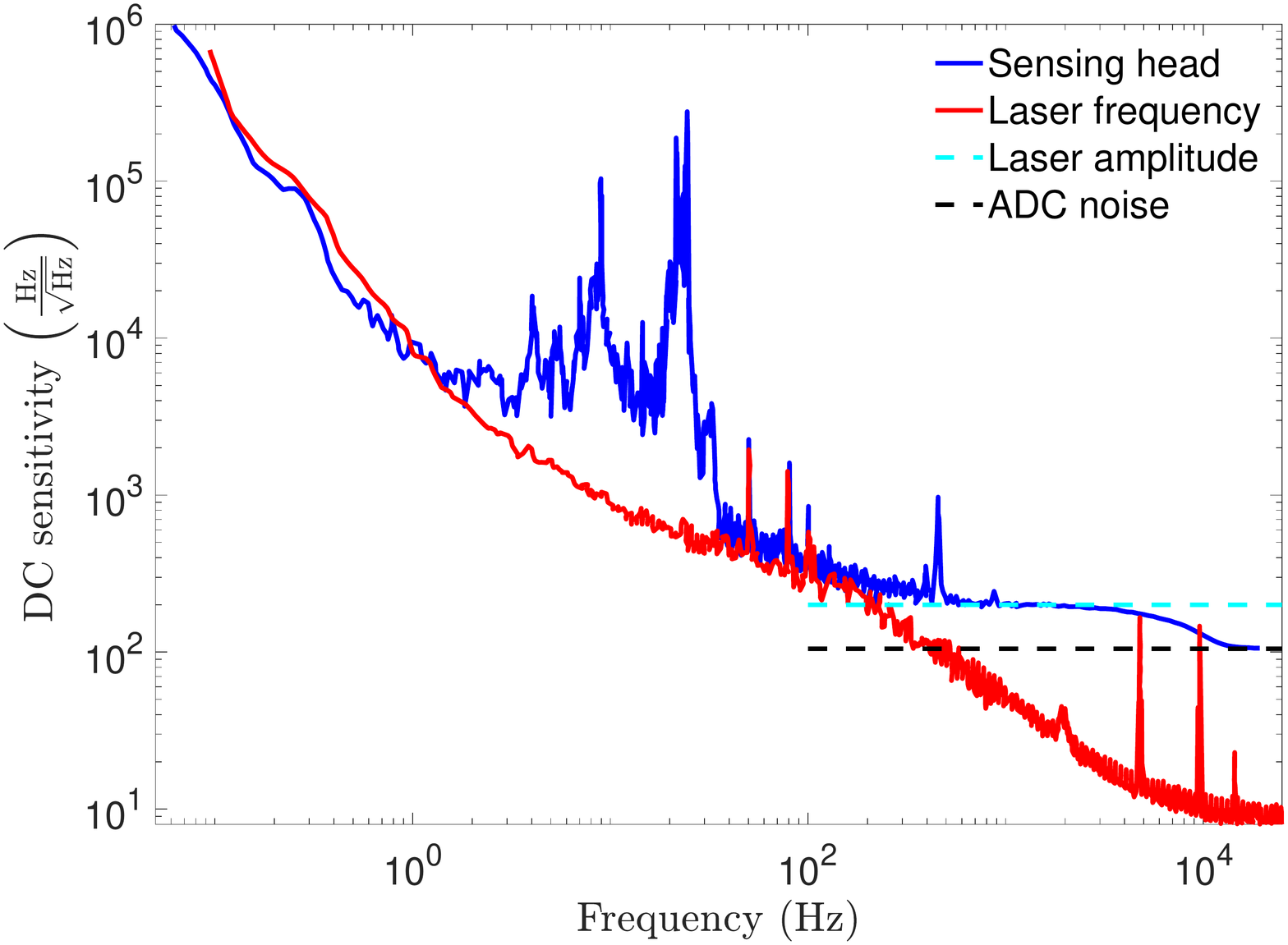}
\phantomcaption
\label{subfig:dcsens}
\end{subfigure}
\hfill
\begin{subfigure}{0.48\linewidth}
\begin{flushleft}\textbf{(b)}\end{flushleft}
\vspace{-3mm}
\includegraphics[width=\linewidth]{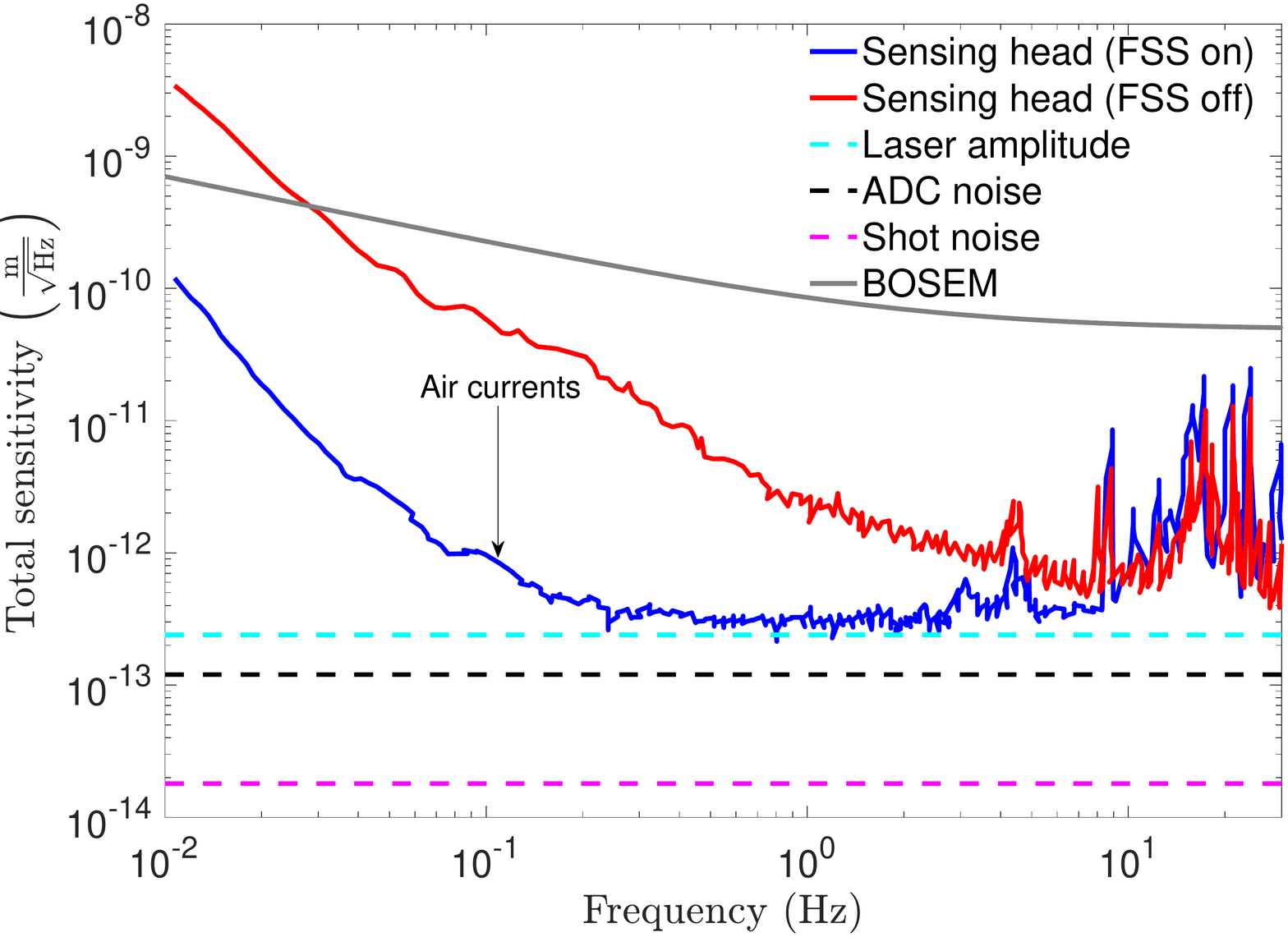}
\phantomcaption
\label{subfig:noise}
\end{subfigure}
\caption{Panel (a) shows the DC sensitivity of the sensing heads without the modulation-demodulation process (calibration factor is $2.85 \times 10^{15}$\,Hz/m in the setup). The DC sensitivity is compared with the laser frequency noise that was independently measured by beating two identical lasers. The two curves are consistent at low frequencies (below 1\,Hz). The dashed lines represent the inferred noise levels from the two largest noise sources near the modulation frequency: laser amplitude noise and ADC noise. The former noise is suppressed above 7\,kHz by the anti-aliasing filter. The laser frequency noise is below the level of $100\,{\rm Hz}/\sqrt{\rm Hz}$ and does not limit the spectrum around the modulation frequency. Panel (b) shows the total sensitivity of the sensing head with the heterodyne readout and frequency noise stabilisation (FSS). The FSS-on curve represents the extend of the sensitivity increase we achieve with the frequency stabilisation. The curve is limited by the residual air currents below 100\,mHz. The dashed lines represent the noise limits of several prominent, incoherent noise sources and show that our new noise floor is consistent with the laser amplitude noise limit at 100\,mHz - 10\,Hz. The BOSEM sensitivity curve~\cite{Bosems} allows for a comparison between the sensitivity of the sensing head and contemporary sensing devices.}
\label{fig:sens}
\end{figure*}

The scheme for achieving our measurement is shown in Fig.~\ref{fig:layout}. A single laser is connected to two separate sensing heads mounted on separate base plates. Each sensing head comprises a simple Michelson interferometer with an internal, fixed-length reference arm and an open port, which forms the external measurement arm. Each measurement beam is then reflected from a fixed mirror mounted on the respective base plate and aligned to maximise the fringe contrast. The interference pattern is detected by a photodiode with a responsivity of 0.9\,A/W. It is mounted on the output of a fiber circulator which connects together the output of the laser, the sensing head and the photodiode. The signal is passed through a 68\,k$\mathrm{\Omega}$ transimpedance amplifier followed by a 7\,kHz low-pass anti-aliasing filter. It is then digitised at a sampling rate of 64\,kHz by a 20\,V peak-to-peak analogue-to-digital converter (ADC). The signal is down-sampled to 32\,kHz once digitised for processing.

We modulate the laser frequency at $\omega_m = 2\pi \times 700$\,Hz and amplitude of $A_m = 0.8$\,GHz with an internal PZT actuator. The power observed at the photodetector is given by the equation~
\cite{DFM}
\begin{equation}
    P(t) = A[1 + C\cos(\phi(t) + m\cos(\omega_m t))],
    \label{eq:power}
\end{equation}
where $A$ is the power amplitude, $C$ is the fringe contrast, $m = 4 \pi A_m \Delta L / c$ is the unit-less modulation index, $\Delta L$ is the longitudinal imbalance of the Michelson arms and $c$ is the speed of light. The variable arm phase, $\phi$, is related to the test mass $x$ displacement and laser wavelength, $\lambda$, via $\phi = 4 \pi x / \lambda$. The signal in Eq.~\ref{eq:power} can be expressed in terms of its Fourier series expansion \cite{DFM}:
\begin{equation}
    P(t) = P_0 + \sum_{n = 1}^\infty 2 CA J_n(m) \cos(\phi + n\pi/2) \cos(n\omega_m t),
    \label{eq:series}
\end{equation}
where $J_n$ is the $n^{\text{th}}$ order Bessel function of the first kind and $P_0 = A(1 + CJ_0(m)\cos(\phi))$. The sum in Eq.~\ref{eq:series} shows that the input modulation at $\omega_m$ manifests in the output spectrum at $\omega_m$ and all higher order harmonics. Furthermore, displacements of the test mass become up-converted to sidebands around each harmonic of $\omega_m$.

The measured signal passes through a transimpedance amplifier and the raw signal is then digitised. The procedure gives us access to the raw power signal (up to the input anti-aliasing filter's cutoff frequency at 7\,kHz) as given in Eq~\ref{eq:power}. The extraction of the signal from the modulation harmonics relies on accurate demodulation, which in turn relies on a local oscillator that corresponds to the modulation source with high fidelity. The key advantage of the digital scheme is that the local oscillator is simultaneously used for the signal demodulation and laser modulation, which leads to a form of digital homodyne readout. This compensates for timing jitter that could otherwise introduce a discrepancy between the modulation source and the local oscillator.


This signal is then demodulated at multiple harmonics of the input modulation frequency
by multiplying the raw input signal by $\sin(k \omega_m t + \psi_k)$ and $\cos(k \omega_m t + \psi_k)$ for $k \in \{1, 2, 3\}$. The demodulation phase, $\psi_k$, represents the phase difference between the modulation signal at its generation and at the point of digitisation together with any time-varying fluctuations such as jitter noise. We tune $\psi_k$ to maximise the signal in the cosine quadrature and reduce the timing fluctuations to a second order effect. The output is given by the equation
\begin{equation}
    S_k = CA J_k(m) \cos(\phi + k\pi/2)
    \label{eq:demod}
\end{equation}

We take the ratio of two harmonics to construct our elliptical Lissajous figure. We derive the arm phase for the first sensing head from the ratio of the first two signals, which is given by
\begin{equation}
    \frac{S_1}{S_2} = \frac{J_1(m)}{J_2(m)} \tan(\phi).
    \label{eq:ratio}
\end{equation}
To simplify the process further, we tune the modulation depth, $A_m$, to achieve the ratio of the Bessel functions in Eq.~\ref{eq:ratio} equal to unity (circular Lissajous figure). This leads to improvements in linearity of the measurement as well as preventing the introduction of non-stationary noise. Since we can tune $A_m$ only for one sensing head, we tuned the length between the second sensing head and the measured mirror to achieve the ratio of the Bessel functions for the second and third harmonics of the sensing head close to unity. The arm phase is extracted in real time through Eq.~\ref{eq:ratio}. Through digital demodulation, we obtain signals $S_1$ and $S_2$ in real time. We then apply the four-quadrant arctangent function to this pair of values per sample on the assumption that the Lissajous figure is circular. Fluctuations and drifts in the parameters of the Lissajous figure introduce non-linear effects in the real-time phase readout. We analysed these effects by applying an ellipse fitting algorithm in post-processing to compare to the performance of the real-time phase extraction scheme. This comparison is addressed in Section~\ref{subsec:linearity}. A more complex real-time phase extraction scheme that does not rely on fixed values of the ellipse parameters could later be adopted if more sensing heads need to be probed with various longitudinal imbalances or drifts become an issue~\cite{DFM}.

\section{\label{sec:sens} Experimental results}

The sensitivity analysis of our interferometric sensing heads was performed on a fixed test mass in order to effectively suppress the presence of ground motion and `zero out' the displacement signal. The observed sensitivity to displacement is shown in Fig.~\ref{subfig:noise}. The achieved sensitivity is $3\times 10^{-13} \text{m} / \sqrt{\text{Hz}}$ at 1\,Hz and is a factor of 300 better compared to the BOSEM, one of the currently used sensors in \textit{Advanced LIGO}~\cite{Bosems}. In this section, we discuss the noise analysis of the measured sensitivity.

\subsection{Environmental disturbances}

We mitigated the perturbation of the laser beams via air currents, which otherwise form the dominant noise source at frequencies below 1\,Hz. We suppressed these effects by placing the components into a sealed, foam-padded box. The box has reduced the air currents by three orders of magnitude around 1\,Hz. However, the noise is still seen below 100\,mHz in Fig.~\ref{subfig:noise}. At these frequencies, our insulation becomes less efficient due to the air pressure fluctuations in the laboratory. We expect that in-vacuum operation of the sensing heads will further suppress the noise below 100\,mHz.

Seismic noise dominates the measured spectrum at 10--100\,Hz because the common-mode-rejection of our setup vanishes above 10\,Hz. The sensing heads do not share the same base plate and are physically separated by 20\,cm. Above 100\,Hz, our passive isolation system suppresses the coupling of the ground vibrations. We achieve the isolation by positioning the sensing heads on metal plates supported by rubber pads. The sealed box is also supported by another set of rubber pads.

\subsection{Laser frequency noise}

Our displacement sensitivity is strongly limited by the laser frequency noise below 10\,Hz as shown in Fig.~\ref{subfig:dcsens}. The figure also shows the readout signal calibrated to the units of Hz/$\sqrt{\rm Hz}$ when no modulation is provided to the laser and the Michelson interferometer is in the gray state: the readout power is at half maximum. In this state, the detector is sensitive to the displacement noise of the test mass, laser frequency and intensity noise as well as the readout electronics.

The frequency noise rises sharply towards lower frequencies and causes significant drifts in the displacement measurement over long periods of time. It is the most significant noise source in our region of interest and poses a significant limit to improvements beyond existing displacement sensor performance. We thus devoted our attention to suppressing this noise to improve the performance of the sensor in this key region.

Typically, the laser sensing noise in Michelson interferometers is suppressed by reducing the longitudinal imbalance between the two arms ($\Delta L = 0$\,m). However, our scheme intentionally uses the imbalance to read out the signal at multiple harmonics of the modulation frequency, $\omega_m$. Therefore, instead of reducing the imbalance, we follow the approach from~\cite{Gerberding_2017, Gerberding_2021} and feed the reference head readout channel back to the laser modulation servo as shown in Fig.~\ref{fig:layout}. This has the effect of rejecting signal, common to both sensing heads, from the second sensing head (the `measurement' head). The displacement spectrum obtained in closed-loop operation is shown in Fig.~\ref{subfig:noise}. Any true motion of the test mass that is detected by the measurement head will be independent of the reference head readout and thus will not be suppressed by the frequency stabilisation servo. Therefore, the noise suppression achieved in Fig.~\ref{subfig:noise} represents a true improvement in the signal-to-noise ratio.


\subsection{Down-converted noises}

Aside from noises that naturally sit at the frequencies measured by the sensing head, there are noise sources which are shifted in frequency according to the choice of modulation frequency. The former noise types, such as laser frequency noise, manifest in the same way as the displacement signal, $\phi$, in Eq.~\ref{eq:power}. Therefore, they undergo the same shift from sidebands around DC up to sidebands around $\omega_m$ and then shift back to DC upon demodulation. The latter types of noise, such as analogue-to-digital converter noise, laser intensity noise, transimpedance amplifier noise, and shot noise, couple directly into the series terms in Eq.~\ref{eq:series}. These noises, therefore, only undergo the down-conversion from the demodulation, which leads to displacement noise coupling of $n_x(\Omega) \propto n_i(\omega_m \pm \Omega)$ for a source $i$ of this type.

This important distinction is a key practical consideration in our choice of modulation frequency. The region around the modulation frequency and its harmonics must be intrinsically lower in noise compared to the signal in the region of interest. Otherwise, the down-converted noise will dominate over the signal in the readout channel. For this very reason, we avoided placing the modulation frequency in the intermediate frequencies dominated by the forest of mechanical resonances. We choose the modulation frequency of 700\,Hz because the laser frequency noise and seismic disturbances are below the broadband laser amplitude noise at this frequency as shown in Fig.~\ref{subfig:dcsens}. In the unconstrained mirror application, such as in the readout of a multiple-degree-of-freedom seismometer~\cite{Compact6D}, this factor will be of particular importance as the seismic measurement will become dominant over other noise sources and will itself be down-converted and potentially obscure the measurement at other frequencies. We are interested in enhancing the precision towards the low frequency limit and, fortunately, the ground motion inherently decreases towards the high frequencies.
Thus, if we place the modulation at an appropriately high frequency, the down-converted ground motion should not pose an issue.

\begin{figure*}[t]
\begin{subfigure}{0.48\linewidth}
\begin{flushleft}\textbf{(a)}\end{flushleft}
\vspace{-3mm}
\includegraphics[width=\linewidth]{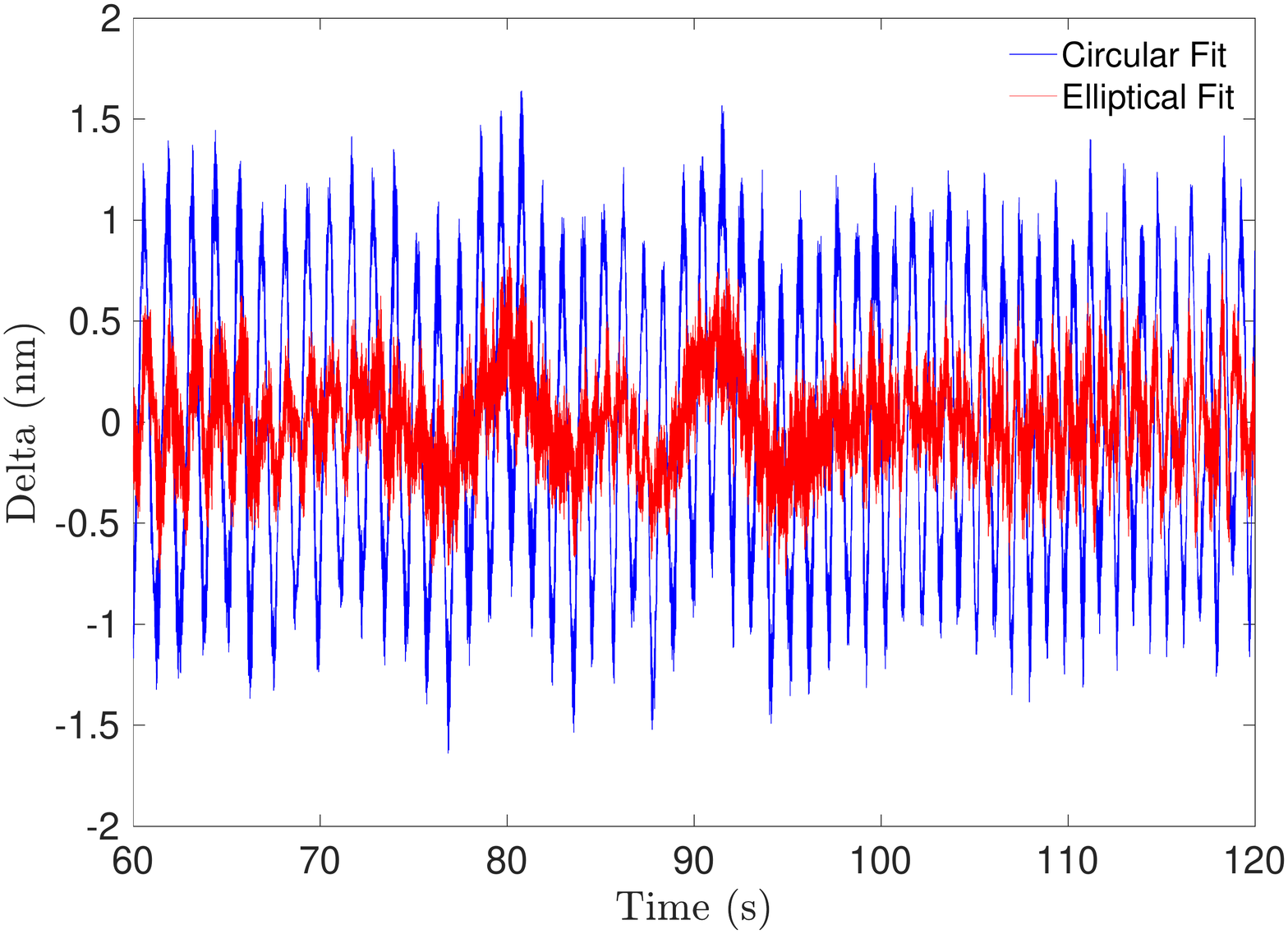}
\phantomcaption
\label{subfig:nltime}
\end{subfigure}
\hfill
\begin{subfigure}{0.48\linewidth}
\begin{flushleft}\textbf{(b)}\end{flushleft}
\vspace{-3mm}
\includegraphics[width=\linewidth]{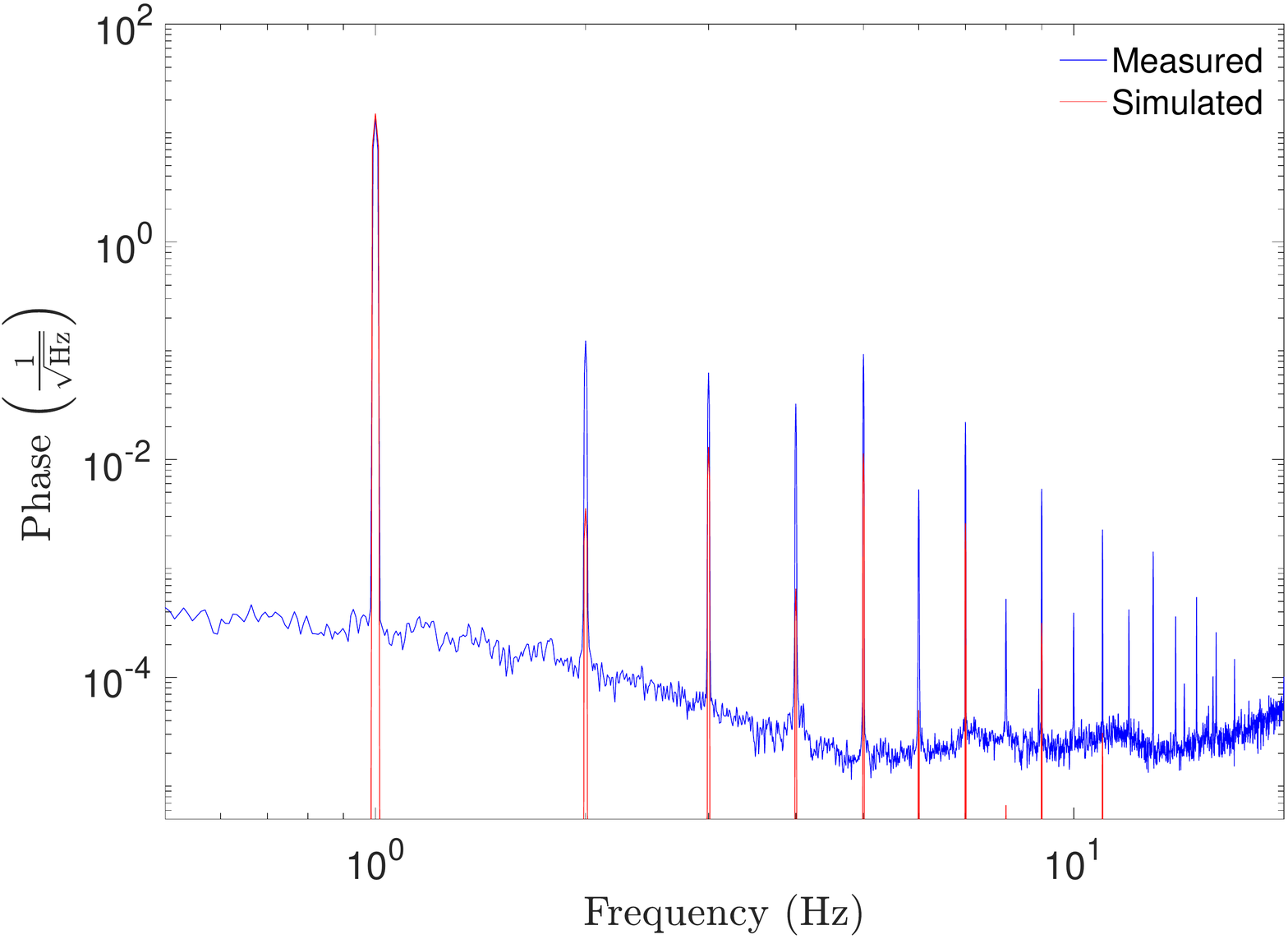}
\phantomcaption
\label{subfig:nlfreq}
\end{subfigure}

\caption{Panel (a) shows the residual displacement from non-linear coupling observed when performing a sweep over the laser wavelength using temperature tuning. The circular fit represents a phase extraction algorithm that assumes balanced harmonics. The elliptical fit shows the improved linearity that we obtained through a more advanced algorithm that applies ellipse fitting to the Lissajous figure. Panel (b) shows the phase spectrum of a 1\,Hz excitation of the laser wavelength using the PZT. The measured data represents the upper limit for the distribution of signal to higher harmonics due to non-linear coupling. The simulated curve represents our estimate of the spectrum that would correspond to the elliptical fit in panel (a).}
\label{fig:non_linear}
\end{figure*}

\subsection{Linearity} \label{subsec:linearity}
Interferometric devices have an exceedingly narrow span of linearity in their raw readout (less than a quarter-wavelength).
The long-range linearity is recovered via the construction of a Lissajous figure using the aforementioned dual-quadrature readout scheme. This technique relies on the correct determination of the centre position and eccentricity of the Lissajous figure in the phase extraction using the four-quadrant arctangent. If these ellipse parameters are determined poorly, the phase signal couples non-linearly into the readout channel, making non-linearities a particular area of concern in the literature~\cite{CompactReview}.

A key source of non-linear coupling we can demonstrate is an imbalance in the amplitude of the quadrature signals, which produces a non-circular Lissajous figure. This can occur through a fluctuation in the modulation depth, $A_m$, which disrupts the equal ratio of the Bessel functions in the first two harmonics ($J_1(m) \neq J_2(m)$ in Eq.~\ref{eq:ratio}). We investigated the long-range linearity of the system by sweeping through the laser wavelength using the laser's temperature control. We swept over a range of 1\,nm corresponding to 62 full rotations of the Lissajous figure. We applied two phase extraction algorithms in post-processing, shown in Fig.~\ref{subfig:nltime}, using a circular and elliptical fit. The former algorithm mimics our real-time phase extraction technique that assumes balanced harmonics, giving a residual RMS displacement of 0.68\,nm over an effective displacement of 48\,\textmugreek m. The latter algorithm applies an elliptical fit to the many-period Lissajous figure constructed from the full sweep in order to correct for non-zero eccentricity and offsets in the centre of the Lissajous figure. Algorithms developed in \cite{EllipseFitting2, EllipseFitting3} can be readily adopted to perform the ellipse fitting. The residual RMS in this case was reduced to 0.25\,nm.

We also approached the investigation of non-linear coupling through another method, consisting of a controlled 1\,Hz excitation of the laser frequency using the laser PZT. The amplitude of the oscillation was set at approximately half the Lissajous period in order to maximise the non-linear effect (we were limited in our choice of amplitude by the range of the PZT). The amplitude spectrum of the phase derived from the real-time readout, shown in Fig.~\ref{subfig:nlfreq}, contains a forest of peaks at multiples of 1\,Hz, which represent the non-linear coupling of the pure 1\,Hz input signal. The measured curve should be treated as an upper limit to the non-linearity of the system, as the non-linear response of the PZT is dominant over the non-linearity of the sensing head readout. We define a measure of the single-period non-linearity as the ratio of the RMS contribution from the non-linear frequency components to the total RMS of the signal. Under this definition, the measured spectrum shows a non-linear measure of 1.2\%. The value is similar to the non-linearity observed in~\cite{Isleif_16} for TLB 6700 Velocity laser from Newport. The simulated curve in Fig.~\ref{subfig:nlfreq} contains the estimated spectrum for an excitation with a non-linear RMS contribution that has been matched to the RMS of the elliptical fit in Fig.~\ref{subfig:nltime}. This model exhibits a non-linear measure of 0.12\%.

In applications that experience larger displacements, such as the position sensing of a suspension, the range of the signals will span multiple interference fringes, making the coupling of non-linearities more prominent than in the short displacement range of our fixed-mirror measurement. Naturally, this makes the topic of linearity particularly relevant.

\section{\label{sec:conc}Conclusion}
Compact, interferometric position sensors are well positioned as candidates for the next generation of displacement readout in precision applications. Two such examples, the 6D seismometer~\cite{Compact6D} and the \textit{Advanced LIGO} quadruple suspensions~\cite{LIGOQUAD}, can benefit from compact interferometers.

We presented a commercially available, interferometric displacement sensor which makes use of the sharp sub-fringe response of a Michelson interferometer. Using deep frequency modulation techniques, we were able to linearise the response over the range of a full fringe and a simple fringe-counting algorithm was implemented to extend the range over multiple fringes. Our fixed mirror measurement showed an improved level of sensitivity over existing shadow sensors. This, together with the sensor's compact design, means it is a suitable candidate for future upgrades and can be readily implemented with little impact on existing infrastructure. The complexity has been effectively shifted to the digital domain. This does make the sensors more difficult to implement in situations where existing computational systems are lacking, but in the case of current GW detectors, this will not be an issue.

We were able to overcome the most significant initial sensitivity limitation, the coupling of laser frequency noise, through the use of one sensing head as a frequency reference. Particularly in multiple degree-of-freedom readout situations, where all sensing heads can receive input from the same laser source, this is a very low-impact solution to improving upon current displacement sensing methods.
In a future upgrade, we will read out the signal from the anti-symmetric port of the Michelson interferometer. The approach will help suppress extra noise from the fiber between the sensing head and the photodetector~\cite{Fleddermann_2018}.

Finally, we have discussed the occurrence of non-linearities in the presence of phase extraction algorithms that apply circular and fitted, static ellipse parameters. The application of our displacement sensor in, for example, suspended test masses may require the routine correction of the parameters of the Lissajous fit. This may arise due to long-term effects such as drifts in the modulation depth and timing jitter, but also due to a transformation of the Lissajous figure over high-RMS displacements, which will vary significantly depending on the specifics of the application.

A more detailed study on the impact of drifts in the parameters of the elliptical fit will be left for future work. Our simplified real-time phase extraction technique may not be suitable for high-RMS applications such as the displacement sensing of a suspended mass with no seismic pre-isolation. In this case, however, it is possible to employ a more advanced, dynamic algorithm that can apply parameter corrections in real-time and maintain linearity more robustly~\cite{DFM}. We are in the process of testing the effectiveness of such an algorithm applied to the sensing heads presented.

The impact of non-linearities in the phase readout will be suppressed in applications such as the \textit{Advanced LIGO} quadruple suspensions. This is due to their significantly lower RMS displacement, which is achieved by the previous stages of seismic isolation~\cite{LIGOSeismicStrategy}. In this case, the RMS displacement should be lower than that simulated by our frequency sweep in Section~\ref{subsec:linearity}, which makes our set-up suitable for such an application already, without significant expansion of the phase extraction scheme.

\section*{Acknowledgements}
We thank members of the LIGO SWG group for useful discussions and Michael Ross for his valuable feedback.
The authors acknowledge the support of the Institute for Gravitational Wave Astronomy at the University of Birmingham, STFC 2018 Equipment Call ST/S002154/1, STFC 'Astrophysics at the University of Birmingham' grant ST/S000305/1. A.S.U. and J.S. are supported by STFC studentships 2117289 and 2116965. O.G. is supported by the Deutsche Forschungsgemeinschaft (DFG, German Research Foundation) under Germany's Excellence Strategy---EXC 2121 ``Quantum Universe''---390833306, and by the German Federal Ministry of Education and Research (BMBF, Project 05A20GU5).

\bibliographystyle{unsrtnat}
\bibliography{main}

\begin{thebibliography}{37}
\providecommand{\natexlab}[1]{#1}
\providecommand{\url}[1]{\texttt{#1}}
\expandafter\ifx\csname urlstyle\endcsname\relax
  \providecommand{\doi}[1]{doi: #1}\else
  \providecommand{\doi}{doi: \begingroup \urlstyle{rm}\Url}\fi

\bibitem[Abbott et~al.(2016)Abbott, Abbott, Abbott, Abernathy, Acernese,
  Ackley, Adams, Adams, Addesso, Adhikari, et~al.]{BBHDetection}
B.~P. Abbott, R.~Abbott, T.~D. Abbott, M.~R. Abernathy, F.~Acernese, K.~Ackley,
  C.~Adams, T.~Adams, P.~Addesso, R.~X. Adhikari, et~al.
\newblock Observation of gravitational waves from a binary black hole merger.
\newblock \emph{Phys. Rev. Lett.}, 116:\penalty0 061102, Feb 2016.
\newblock \doi{10.1103/PhysRevLett.116.061102}.
\newblock URL \url{https://link.aps.org/doi/10.1103/PhysRevLett.116.061102}.

\bibitem[Abbott et~al.(2019)Abbott, Abbott, Abbott, Abraham, Acernese, Ackley,
  Adams, Adhikari, Adya, Affeldt, et~al.]{Catalogue1}
B.~P. Abbott, R.~Abbott, T.~D. Abbott, S.~Abraham, F.~Acernese, K.~Ackley,
  C.~Adams, R.~X. Adhikari, V.~B. Adya, C.~Affeldt, et~al.
\newblock Gwtc-1: A gravitational-wave transient catalog of compact binary
  mergers observed by ligo and virgo during the first and second observing
  runs.
\newblock \emph{Phys. Rev. X}, 9:\penalty0 031040, Sep 2019.
\newblock \doi{10.1103/PhysRevX.9.031040}.
\newblock URL \url{https://link.aps.org/doi/10.1103/PhysRevX.9.031040}.

\bibitem[Abbott et~al.(2021)Abbott, Abbott, Abraham, Acernese, Ackley, Adams,
  Adams, Adhikari, Adya, Affeldt, et~al.]{Catalogue2}
R.~Abbott, T.~D. Abbott, S.~Abraham, F.~Acernese, K.~Ackley, A.~Adams,
  C.~Adams, R.~X. Adhikari, V.~B. Adya, C.~Affeldt, et~al.
\newblock Gwtc-2: Compact binary coalescences observed by ligo and virgo during
  the first half of the third observing run.
\newblock \emph{Phys. Rev. X}, 11:\penalty0 021053, Jun 2021.
\newblock \doi{10.1103/PhysRevX.11.021053}.
\newblock URL \url{https://link.aps.org/doi/10.1103/PhysRevX.11.021053}.

\bibitem[Aasi et~al.(2015)Aasi, Abbott, Abbott, Abbott, Abernathy, Ackley,
  Adams, Adams, Addesso, Adhikari, et~al.]{AdvLIGO}
J~Aasi, B~P Abbott, R~Abbott, T~Abbott, M~R Abernathy, K~Ackley, C~Adams,
  T~Adams, P~Addesso, R~X Adhikari, et~al.
\newblock Advanced {LIGO}.
\newblock \emph{Classical and Quantum Gravity}, 32\penalty0 (7):\penalty0
  074001, mar 2015.
\newblock \doi{10.1088/0264-9381/32/7/074001}.
\newblock URL \url{https://doi.org/10.1088/0264-9381/32/7/074001}.

\bibitem[Acernese et~al.(2014)Acernese, Agathos, Agatsuma, Aisa, Allemandou,
  Allocca, Amarni, Astone, Balestri, Ballardin, et~al.]{AdvVirgo}
F~Acernese, M~Agathos, K~Agatsuma, D~Aisa, N~Allemandou, A~Allocca, J~Amarni,
  P~Astone, G~Balestri, G~Ballardin, et~al.
\newblock Advanced virgo: a second-generation interferometric gravitational
  wave detector.
\newblock \emph{Classical and Quantum Gravity}, 32\penalty0 (2):\penalty0
  024001, dec 2014.
\newblock \doi{10.1088/0264-9381/32/2/024001}.
\newblock URL \url{https://doi.org/10.1088/0264-9381/32/2/024001}.

\bibitem[Buikema et~al.(2020)Buikema, Cahillane, Mansell, Blair, Abbott, Adams,
  Adhikari, Ananyeva, Appert, Arai, et~al.]{Buikema2020}
A.~Buikema, C.~Cahillane, G.~L. Mansell, C.~D. Blair, R.~Abbott, C.~Adams,
  R.~X. Adhikari, A.~Ananyeva, S.~Appert, K.~Arai, et~al.
\newblock Sensitivity and performance of the advanced ligo detectors in the
  third observing run.
\newblock \emph{Phys. Rev. D}, 102:\penalty0 062003, Sep 2020.
\newblock \doi{10.1103/PhysRevD.102.062003}.
\newblock URL \url{https://link.aps.org/doi/10.1103/PhysRevD.102.062003}.

\bibitem[Amaro-Seoane et~al.(2017)Amaro-Seoane, Audley, Babak, Baker, Barausse,
  Bender, Berti, Binetruy, Born, Bortoluzzi, et~al.]{LISA}
Pau Amaro-Seoane, Heather Audley, Stanislav Babak, John Baker, Enrico Barausse,
  Peter Bender, Emanuele Berti, Pierre Binetruy, Michael Born, Daniele
  Bortoluzzi, et~al.
\newblock Laser interferometer space antenna.
\newblock \emph{ArXiv}, 2017.

\bibitem[Carbone et~al.(2012)Carbone, Aston, Cutler, Freise, Greenhalgh,
  Heefner, Hoyland, Lockerbie, Lodhia, Robertson, Speake, Strain, and
  Vecchio]{Bosems}
L~Carbone, S~M Aston, R~M Cutler, A~Freise, J~Greenhalgh, J~Heefner, D~Hoyland,
  N~A Lockerbie, D~Lodhia, N~A Robertson, C~C Speake, K~A Strain, and
  A~Vecchio.
\newblock Sensors and actuators for the advanced {LIGO} mirror suspensions.
\newblock \emph{Classical and Quantum Gravity}, 29\penalty0 (11):\penalty0
  115005, may 2012.
\newblock \doi{10.1088/0264-9381/29/11/115005}.
\newblock URL \url{https://doi.org/10.1088/0264-9381/29/11/115005}.

\bibitem[Magee et~al.(2021)Magee, Chatterjee, Singer, Sachdev, Kovalam, Mo,
  Anderson, Brady, Brockill, Cannon, et~al.]{EarlyWarning}
Ryan Magee, Deep Chatterjee, Leo~P. Singer, Surabhi Sachdev, Manoj Kovalam,
  Geoffrey Mo, Stuart Anderson, Patrick Brady, Patrick Brockill, Kipp Cannon,
  et~al.
\newblock First demonstration of early warning gravitational-wave alerts.
\newblock \emph{The Astrophysical Journal Letters}, 910\penalty0 (2):\penalty0
  L21, apr 2021.
\newblock \doi{10.3847/2041-8213/abed54}.
\newblock URL \url{https://doi.org/10.3847/2041-8213/abed54}.

\bibitem[Branchesi(2016)]{MultiMessenger}
Marica Branchesi.
\newblock Multi-messenger astronomy: gravitational waves, neutrinos, photons,
  and cosmic rays.
\newblock \emph{Journal of Physics: Conference Series}, 718:\penalty0 022004,
  may 2016.
\newblock \doi{10.1088/1742-6596/718/2/022004}.
\newblock URL \url{https://doi.org/10.1088/1742-6596/718/2/022004}.

\bibitem[Amaro-Seoane et~al.(2007)Amaro-Seoane, Gair, Freitag, Miller, Mandel,
  Cutler, and Babak]{IMBHDetection}
Pau Amaro-Seoane, Jonathan~R Gair, Marc Freitag, M~Coleman Miller, Ilya Mandel,
  Curt~J Cutler, and Stanislav Babak.
\newblock Intermediate and extreme mass-ratio
  inspirals{\textemdash}astrophysics, science applications and detection using
  {LISA}.
\newblock \emph{Classical and Quantum Gravity}, 24\penalty0 (17):\penalty0
  R113--R169, aug 2007.
\newblock \doi{10.1088/0264-9381/24/17/r01}.
\newblock URL \url{https://doi.org/10.1088/0264-9381/24/17/r01}.

\bibitem[Yu et~al.(2018)Yu, Martynov, Vitale, Evans, Shoemaker, Barr, Hammond,
  Hild, Hough, Huttner, Rowan, Sorazu, Carbone, Freise, Mow-Lowry, Dooley,
  Fulda, Grote, and Sigg]{5Hz}
Hang Yu, Denis Martynov, Salvatore Vitale, Matthew Evans, David Shoemaker,
  Bryan Barr, Giles Hammond, Stefan Hild, James Hough, Sabina Huttner, Sheila
  Rowan, Borja Sorazu, Ludovico Carbone, Andreas Freise, Conor Mow-Lowry,
  Katherine~L. Dooley, Paul Fulda, Hartmut Grote, and Daniel Sigg.
\newblock Prospects for detecting gravitational waves at 5 hz with ground-based
  detectors.
\newblock \emph{Phys. Rev. Lett.}, 120:\penalty0 141102, Apr 2018.
\newblock \doi{10.1103/PhysRevLett.120.141102}.
\newblock URL \url{https://link.aps.org/doi/10.1103/PhysRevLett.120.141102}.

\bibitem[Ubhi et~al.(2021)Ubhi, Smetana, Zhang, Cooper, Prokhorov, Bryant,
  Hoyland, Miao, and Martynov]{Compact6D}
Amit~Singh Ubhi, Jiri Smetana, Teng Zhang, Sam Cooper, Leonid Prokhorov, John
  Bryant, David Hoyland, Haixing Miao, and Denis Martynov.
\newblock A six degree-of-freedom fused silica seismometer: design~and tests of
  a metal prototype.
\newblock \emph{Classical and Quantum Gravity}, 39\penalty0 (1):\penalty0
  015006, dec 2021.
\newblock \doi{10.1088/1361-6382/ac39b9}.
\newblock URL \url{https://doi.org/10.1088/1361-6382/ac39b9}.

\bibitem[Venkateswara et~al.(2014)Venkateswara, Hagedorn, Turner, Arp, and
  Gundlach]{BRS_2014}
Krishna Venkateswara, Charles~A. Hagedorn, Matthew~D. Turner, Trevor Arp, and
  Jens~H. Gundlach.
\newblock A high-precision mechanical absolute-rotation sensor.
\newblock \emph{Review of Scientific Instruments}, 85\penalty0 (1):\penalty0
  015005, 2014.
\newblock \doi{10.1063/1.4862816}.
\newblock URL \url{https://doi.org/10.1063/1.4862816}.

\bibitem[Ross et~al.(2021)Ross, Venkateswara, Hagedorn, Leupold, Forsyth,
  Wegner, Shaw, Lee, and Gundlach]{TorsionMichelson}
M.~P. Ross, K.~Venkateswara, C.~A. Hagedorn, C.~J. Leupold, P.~W.~F. Forsyth,
  J.~D. Wegner, E.~A. Shaw, J.~G. Lee, and J.~H. Gundlach.
\newblock A low-frequency torsion pendulum with interferometric readout.
\newblock \emph{Review of Scientific Instruments}, 92\penalty0 (5):\penalty0
  054502, 2021.
\newblock \doi{10.1063/5.0043098}.
\newblock URL \url{https://doi.org/10.1063/5.0043098}.

\bibitem[Staley et~al.(2014)Staley, Martynov, Abbott, Adhikari, Arai, Ballmer,
  Barsotti, Brooks, DeRosa, Dwyer, Effler, Evans, Fritschel, Frolov, Gray,
  Guido, Gustafson, Heintze, Hoak, Izumi, Kawabe, King, Kissel, Kokeyama,
  Landry, McClelland, Miller, Mullavey, O'Reilly, Rollins, Sanders, Schofield,
  Sigg, Slagmolen, Smith-Lefebvre, Vajente, Ward, and Wipf]{Staley_2014}
A~Staley, D~Martynov, R~Abbott, R~X Adhikari, K~Arai, S~Ballmer, L~Barsotti,
  A~F Brooks, R~T DeRosa, S~Dwyer, A~Effler, M~Evans, P~Fritschel, V~V Frolov,
  C~Gray, C~J Guido, R~Gustafson, M~Heintze, D~Hoak, K~Izumi, K~Kawabe, E~J
  King, J~S Kissel, K~Kokeyama, M~Landry, D~E McClelland, J~Miller, A~Mullavey,
  B~O'Reilly, J~G Rollins, J~R Sanders, R~M~S Schofield, D~Sigg, B~J~J
  Slagmolen, N~D Smith-Lefebvre, G~Vajente, R~L Ward, and C~Wipf.
\newblock Achieving resonance in the advanced {LIGO} gravitational-wave
  interferometer.
\newblock \emph{Classical and Quantum Gravity}, 31\penalty0 (24):\penalty0
  245010, nov 2014.
\newblock \doi{10.1088/0264-9381/31/24/245010}.
\newblock URL \url{https://doi.org/10.1088/0264-9381/31/24/245010}.

\bibitem[Shaw et~al.(2022)Shaw, Ross, Hagedorn, Adelberger, and
  Gundlach]{TorsionDarkMatter}
E.~A. Shaw, M.~P. Ross, C.~A. Hagedorn, E.~G. Adelberger, and J.~H. Gundlach.
\newblock Torsion-balance search for ultralow-mass bosonic dark matter.
\newblock \emph{Phys. Rev. D}, 105:\penalty0 042007, Feb 2022.
\newblock \doi{10.1103/PhysRevD.105.042007}.
\newblock URL \url{https://link.aps.org/doi/10.1103/PhysRevD.105.042007}.

\bibitem[Thomas et~al.(2019)Thomas, Bredendiek, and
  Pohl]{CompactReferenceSensor}
Sven Thomas, Christian Bredendiek, and Nils Pohl.
\newblock A sige-based 240-ghz fmcw radar system for high-resolution
  measurements.
\newblock \emph{IEEE Transactions on Microwave Theory and Techniques},
  67\penalty0 (11):\penalty0 4599--4609, 2019.
\newblock \doi{10.1109/TMTT.2019.2916851}.

\bibitem[Yu et~al.(2019)Yu, Pfeiffer, Morsali, Yang, and
  Fontaine]{AbsoluteDistanceSensor}
Wenhui Yu, Pierre Pfeiffer, Alireza Morsali, Jianming Yang, and Jo\"{e}l
  Fontaine.
\newblock Comb-calibrated frequency sweeping interferometry for absolute
  distance and vibration measurement.
\newblock \emph{Opt. Lett.}, 44\penalty0 (20):\penalty0 5069--5072, Oct 2019.
\newblock \doi{10.1364/OL.44.005069}.
\newblock URL \url{http://opg.optica.org/ol/abstract.cfm?URI=ol-44-20-5069}.

\bibitem[Sosin et~al.(2019)Sosin, Mainaud-Durand, Rude, and
  Rutkowski]{LHCAlignment}
M.~Sosin, H.~Mainaud-Durand, V.~Rude, and J.~Rutkowski.
\newblock {Frequency sweeping interferometry for robust and reliable distance
  measurements in harsh accelerator environment}.
\newblock In Erik Novak and James~D. Trolinger, editors, \emph{Applied Optical
  Metrology III}, volume 11102, pages 145 -- 161. International Society for
  Optics and Photonics, SPIE, 2019.
\newblock \doi{10.1117/12.2529157}.
\newblock URL \url{https://doi.org/10.1117/12.2529157}.

\bibitem[Watchi et~al.(2018)Watchi, Cooper, Ding, Mow-Lowry, and
  Collette]{CompactReview}
Jennifer Watchi, Sam Cooper, Binlei Ding, Conor~M. Mow-Lowry, and Christophe
  Collette.
\newblock Contributed review: A review of compact interferometers.
\newblock \emph{Review of Scientific Instruments}, 89\penalty0 (12):\penalty0
  121501, 2018.
\newblock \doi{10.1063/1.5052042}.
\newblock URL \url{https://doi.org/10.1063/1.5052042}.

\bibitem[de~la Rue et~al.(1972)de~la Rue, Humphryes, Mason, and
  Ash]{Compact1972}
R.M. de~la Rue, R.F. Humphryes, I.M. Mason, and E.A. Ash.
\newblock Acoustic-surface-wave amplitude and phase measurements using laser
  probes.
\newblock \emph{Proceedings of the Institution of Electrical Engineers},
  119:\penalty0 117--126(9), February 1972.
\newblock ISSN 0020-3270.
\newblock URL
  \url{https://digital-library.theiet.org/content/journals/10.1049/piee.1972.0021}.

\bibitem[Weichert et~al.(2012)Weichert, Köchert, Köning, Flügge, Andreas,
  Kuetgens, and Yacoot]{CompactLinear}
C~Weichert, P~Köchert, R~Köning, J~Flügge, B~Andreas, U~Kuetgens, and
  A~Yacoot.
\newblock A heterodyne interferometer with periodic nonlinearities smaller than
  $\pm$10{\hspace{0.167em}}pm.
\newblock \emph{Measurement Science and Technology}, 23\penalty0 (9):\penalty0
  094005, jul 2012.
\newblock \doi{10.1088/0957-0233/23/9/094005}.
\newblock URL \url{https://doi.org/10.1088/0957-0233/23/9/094005}.

\bibitem[Armano et~al.(2016)Armano, Audley, Auger, Baird, Bassan, Binetruy,
  Born, Bortoluzzi, Brandt, Caleno, et~al.]{CompactLISAPF}
M.~Armano, H.~Audley, G.~Auger, J.~T. Baird, M.~Bassan, P.~Binetruy, M.~Born,
  D.~Bortoluzzi, N.~Brandt, M.~Caleno, et~al.
\newblock Sub-femto-$g$ free fall for space-based gravitational wave
  observatories: Lisa pathfinder results.
\newblock \emph{Phys. Rev. Lett.}, 116:\penalty0 231101, Jun 2016.
\newblock \doi{10.1103/PhysRevLett.116.231101}.
\newblock URL \url{https://link.aps.org/doi/10.1103/PhysRevLett.116.231101}.

\bibitem[Robertson et~al.(2002)Robertson, Cagnoli, Crooks, Elliffe, Faller,
  Fritschel, Go{\textdollar}szlig{\textdollar}ler, Grant, Heptonstall, Hough,
  L{\textdollar}uuml{\textdollar}ck, Mittleman, Perreur-Lloyd, Plissi, Rowan,
  Shoemaker, Sneddon, Strain, Torrie, Ward, and Willems]{LIGOQUAD}
N~A Robertson, G~Cagnoli, D~R~M Crooks, E~Elliffe, J~E Faller, P~Fritschel,
  S~Go{\textdollar}szlig{\textdollar}ler, A~Grant, A~Heptonstall, J~Hough,
  H~L{\textdollar}uuml{\textdollar}ck, R~Mittleman, M~Perreur-Lloyd, M~V
  Plissi, S~Rowan, D~H Shoemaker, P~H Sneddon, K~A Strain, C~I Torrie, H~Ward,
  and P~Willems.
\newblock Quadruple suspension design for advanced {LIGO}.
\newblock \emph{Classical and Quantum Gravity}, 19\penalty0 (15):\penalty0
  4043--4058, jul 2002.
\newblock \doi{10.1088/0264-9381/19/15/311}.
\newblock URL \url{https://doi.org/10.1088/0264-9381/19/15/311}.

\bibitem[Royer and Dieulesaint(1986)]{CompactSmall}
Daniel Royer and Eug{\`e}ne Dieulesaint.
\newblock Optical detection of sub-angstrom transient mechanical displacements.
\newblock \emph{IEEE 1986 Ultrasonics Symposium}, pages 527--530, 1986.

\bibitem[Cooper et~al.(2018)Cooper, Collins, Green, Hoyland, Speake, Freise,
  and Mow-Lowry]{HoQITest}
S~J Cooper, C~J Collins, A~C Green, D~Hoyland, C~C Speake, A~Freise, and C~M
  Mow-Lowry.
\newblock A compact, large-range interferometer for precision measurement and
  inertial sensing.
\newblock \emph{Classical and Quantum Gravity}, 35\penalty0 (9):\penalty0
  095007, mar 2018.
\newblock \doi{10.1088/1361-6382/aab2e9}.
\newblock URL \url{https://doi.org/10.1088/1361-6382/aab2e9}.

\bibitem[Cooper et~al.(2022)Cooper, Collins, Prokhorov, Warner, Hoyland, and
  Mow-Lowry]{HoQIGeophone}
S~J Cooper, C~J Collins, L~Prokhorov, J~Warner, D~Hoyland, and C~M Mow-Lowry.
\newblock Interferometric sensing of a commercial geophone.
\newblock \emph{Classical and Quantum Gravity}, 39\penalty0 (7):\penalty0
  075023, mar 2022.
\newblock \doi{10.1088/1361-6382/ac595c}.
\newblock URL \url{https://doi.org/10.1088/1361-6382/ac595c}.

\bibitem[Isleif et~al.(2019)Isleif, Heinzel, Mehmet, and
  Gerberding]{Isleif_2019}
Katharina-Sophie Isleif, Gerhard Heinzel, Moritz Mehmet, and Oliver Gerberding.
\newblock Compact multifringe interferometry with subpicometer precision.
\newblock \emph{Physical Review Applied}, 12\penalty0 (3), Sep 2019.
\newblock ISSN 2331-7019.
\newblock \doi{10.1103/physrevapplied.12.034025}.
\newblock URL \url{http://dx.doi.org/10.1103/PhysRevApplied.12.034025}.

\bibitem[Heinzel et~al.(2010)Heinzel, Cervantes, Mar\'{i}n, Kullmann, Feng, and
  Danzmann]{DFM}
Gerhard Heinzel, Felipe~Guzm\'{a}n Cervantes, Antonio F.~Garc\'{i}a Mar\'{i}n,
  Joachim Kullmann, Wang Feng, and Karsten Danzmann.
\newblock Deep phase modulation interferometry.
\newblock \emph{Opt. Express}, 18\penalty0 (18):\penalty0 19076--19086, Aug
  2010.
\newblock \doi{10.1364/OE.18.019076}.
\newblock URL
  \url{http://www.osapublishing.org/oe/abstract.cfm?URI=oe-18-18-19076}.

\bibitem[Gerberding et~al.(2017)Gerberding, Isleif, Mehmet, Danzmann, and
  Heinzel]{Gerberding_2017}
Oliver Gerberding, Katharina-Sophie Isleif, Moritz Mehmet, Karsten Danzmann,
  and Gerhard Heinzel.
\newblock Laser-frequency stabilization via a quasimonolithic mach-zehnder
  interferometer with arms of unequal length and balanced dc readout.
\newblock \emph{Phys. Rev. Applied}, 7:\penalty0 024027, Feb 2017.
\newblock \doi{10.1103/PhysRevApplied.7.024027}.
\newblock URL \url{https://link.aps.org/doi/10.1103/PhysRevApplied.7.024027}.

\bibitem[Gerberding and Isleif(2021)]{Gerberding_2021}
Oliver Gerberding and Katharina-Sophie Isleif.
\newblock Ghost beam suppression in deep frequency modulation interferometry
  for compact on-axis optical heads.
\newblock \emph{Sensors}, 21\penalty0 (5), 2021.
\newblock ISSN 1424-8220.
\newblock \doi{10.3390/s21051708}.
\newblock URL \url{https://www.mdpi.com/1424-8220/21/5/1708}.

\bibitem[Watkins and Collett(2014)]{EllipseFitting2}
Lionel~R. Watkins and Matthew~J. Collett.
\newblock Ellipse fitting for interferometry. part 2: experimental realization.
\newblock \emph{Appl. Opt.}, 53\penalty0 (32):\penalty0 7697--7703, Nov 2014.
\newblock \doi{10.1364/AO.53.007697}.
\newblock URL \url{http://opg.optica.org/ao/abstract.cfm?URI=ao-53-32-7697}.

\bibitem[Collett and Watkins(2015)]{EllipseFitting3}
M.~J. Collett and L.~R. Watkins.
\newblock Ellipse fitting for interferometry. part 3: dynamic method.
\newblock \emph{J. Opt. Soc. Am. A}, 32\penalty0 (3):\penalty0 491--496, Mar
  2015.
\newblock \doi{10.1364/JOSAA.32.000491}.
\newblock URL
  \url{http://opg.optica.org/josaa/abstract.cfm?URI=josaa-32-3-491}.

\bibitem[Isleif et~al.(2016)Isleif, Gerberding, Schwarze, Mehmet, Heinzel, and
  Cervantes]{Isleif_16}
Katharina-Sophie Isleif, Oliver Gerberding, Thomas~S. Schwarze, Moritz Mehmet,
  Gerhard Heinzel, and Felipe~Guzm\'{a}n Cervantes.
\newblock Experimental demonstration of deep frequency modulation
  interferometry.
\newblock \emph{Opt. Express}, 24\penalty0 (2):\penalty0 1676--1684, Jan 2016.
\newblock \doi{10.1364/OE.24.001676}.
\newblock URL \url{http://opg.optica.org/oe/abstract.cfm?URI=oe-24-2-1676}.

\bibitem[Fleddermann et~al.(2018)Fleddermann, Diekmann, Steier, Tröbs,
  Heinzel, and Danzmann]{Fleddermann_2018}
Roland Fleddermann, Christian Diekmann, Frank Steier, Michael Tröbs, Gerhard
  Heinzel, and Karsten Danzmann.
\newblock
  Sub-pm{\textdollar}$\lbrace$$\lbrace${\textbackslash}sqrt$\lbrace$$\lbrace${\textbackslash}rm
  hz$\rbrace$$\rbrace${\^{}}$\lbrace$-1$\rbrace$$\rbrace$$\rbrace${\textdollar}
  non-reciprocal noise in the {LISA} backlink fiber.
\newblock \emph{Classical and Quantum Gravity}, 35\penalty0 (7):\penalty0
  075007, feb 2018.
\newblock \doi{10.1088/1361-6382/aaa276}.
\newblock URL \url{https://doi.org/10.1088/1361-6382/aaa276}.

\bibitem[Matichard et~al.(2015)Matichard, Lantz, Mittleman, Mason, Kissel,
  Abbott, Biscans, McIver, Abbott, Abbott, Allwine, Barnum, Birch, Celerier,
  Clark, Coyne, DeBra, DeRosa, Evans, Foley, Fritschel, Giaime, Gray, Grabeel,
  Hanson, Hardham, Hillard, Hua, Kucharczyk, Landry, {Le Roux}, Lhuillier,
  Macleod, Macinnis, Mitchell, O'Reilly, Ottaway, Paris, Pele, Puma, Radkins,
  Ramet, Robinson, Ruet, Sarin, Shoemaker, Stein, Thomas, Vargas, Venkateswara,
  Warner, and Wen]{LIGOSeismicStrategy}
F~Matichard, B~Lantz, R~Mittleman, K~Mason, J~Kissel, B~Abbott, S~Biscans,
  J~McIver, R~Abbott, S~Abbott, E~Allwine, S~Barnum, J~Birch, C~Celerier,
  D~Clark, D~Coyne, D~DeBra, R~DeRosa, M~Evans, S~Foley, P~Fritschel, J~A
  Giaime, C~Gray, G~Grabeel, J~Hanson, C~Hardham, M~Hillard, W~Hua,
  C~Kucharczyk, M~Landry, A~{Le Roux}, V~Lhuillier, D~Macleod, M~Macinnis,
  R~Mitchell, B~O'Reilly, D~Ottaway, H~Paris, A~Pele, M~Puma, H~Radkins,
  C~Ramet, M~Robinson, L~Ruet, P~Sarin, D~Shoemaker, A~Stein, J~Thomas,
  M~Vargas, K~Venkateswara, J~Warner, and S~Wen.
\newblock {Seismic isolation of Advanced LIGO: Review of strategy,
  instrumentation and performance}.
\newblock \emph{Classical and Quantum Gravity}, 32\penalty0 (18):\penalty0
  185003, sep 2015.
\newblock ISSN 0264-9381.
\newblock \doi{10.1088/0264-9381/32/18/185003}.
\newblock URL
  \url{http://stacks.iop.org/0264-9381/32/i=18/a=185003?key=crossref.587edd10ec005ab3c52ff8365a2072df}.

\end{thebibliography}

\end{document}